\pdfoutput=1
\documentclass[11pt]{article}

\usepackage[final]{acl}

\usepackage{times}
\usepackage{latexsym}

\usepackage[T1]{fontenc}
\usepackage[utf8]{inputenc}

\usepackage{microtype}

\usepackage{inconsolata}

\usepackage{graphicx}

\usepackage[linesnumbered,ruled]{algorithm2e}
\usepackage{amsmath}
\usepackage{multirow}
\usepackage{pifont}
\usepackage{colortbl}
\usepackage{xspace}

\newcommand{\eg}{\textit{e.g.}\xspace}
\newcommand{\ie}{\textit{i.e.}\xspace}

\usepackage{enumitem}
\usepackage[capitalize]{cleveref}
\crefname{section}{Sec.}{Secs.}
\Crefname{section}{Section}{Sections}
\crefname{table}{Tab.}{Tabs.}
\Crefname{table}{Table}{Tables}

\usepackage{booktabs}
\usepackage{amssymb}
\usepackage{subcaption}

\title{The Security Threat of Compressed Projectors\\in Large Vision-Language Models}

\author{
 \textbf{Yudong Zhang\textsuperscript{1,2}},
 \textbf{Ruobing Xie\textsuperscript{2,\ding{41}}},
 \textbf{Xingwu Sun\textsuperscript{2,4}},
 \textbf{Jiansheng Chen\textsuperscript{3,\ding{41}}},
\\
 \textbf{Zhanhui Kang\textsuperscript{2}},
 \textbf{Di Wang\textsuperscript{2}},
 \textbf{Yu Wang\textsuperscript{1,\ding{41}}}
\\
 \textsuperscript{1}Department of Electronic Engineering, Tsinghua University,\\
 \textsuperscript{2}Large Language Model Department, Tencent,\\
 \textsuperscript{3}School of Computer and Communication Engineering, University of Science and \\ Technology Beijing,
 \textsuperscript{4}Faculty of Science and Technology, University of Macau\\
\small{
 zhangyd16@mails.tsinghua.edu.cn, xrbsnowing@163.com, sunxingwu01@gmail.com, jschen@ustb.edu.cn,
}
\\
\small{
kegokang@tencent.com, diwang@tencent.com, yu-wang@mail.tsinghua.edu.cn. (\ding{41}: Corresponding authors)
}
}

\begin{document}
\maketitle
\begin{abstract}
The choice of a suitable visual language projector (VLP) is critical to the successful training of large visual language models (LVLMs). Mainstream VLPs can be broadly categorized into compressed and uncompressed projectors, and each offers distinct advantages in performance and computational efficiency. However, their security implications have not been thoroughly examined. Our comprehensive evaluation reveals significant differences in their security profiles: compressed projectors exhibit substantial vulnerabilities, allowing adversaries to successfully compromise LVLMs even with minimal knowledge of structure information. In stark contrast, uncompressed projectors demonstrate robust security properties and do not introduce additional vulnerabilities. These findings provide critical guidance for researchers in selecting optimal VLPs that enhance the security and reliability of visual language models. The code is available at \url{https://github.com/btzyd/TCP}.
\end{abstract}

\section{Introduction}
\label{sec:introduction}

Large vision-language models (LVLMs) have achieved remarkable success in various multimedia applications \cite{dai2305instructblip, liu2024visual, bai2025qwen2, wang2024qwen2, chen2024internvl}, particularly in tasks like visual question answering (VQA). A typical LVLM framework consists of three key components: visual encoder (VE), large language model (LLM), and vision-language projector (VLP). Current training methodologies primarily focus on optimizing the VLP to effectively integrate visual features extracted from a pre-trained VE into a pre-trained LLM. This approach substantially minimizes computational demands when compared to building an LVLM from scratch, thereby allowing researchers to leverage existing advancements. Moreover, it ensures greater stability and reliability in both the training process and practical deployment of LVLMs.

Recently, a growing body of research has employed CLIP \cite{radford2021learning, fang2023eva} as the VE and initiated training on various pre-trained LLMs. The choice of VLP plays a crucial role in determining the effectiveness of LVLM training. VLPs can generally be classified into two distinct categories: compressed projectors and uncompressed projectors. \textbf{Compressed projectors}, exemplified by Q-former \cite{li2023blip, dai2305instructblip, zhu2023minigpt}, achieve high computational efficiency by compressing a larger number of visual tokens into a smaller dimension using query tokens. On the other hand, \textbf{uncompressed projectors}, typified by MLP-based architectures \cite{liu2024visual, shi2024eagle}, convert visual tokens to feature dimensions matching the LLM, with the number of output tokens proportional to the number of input tokens.

Previous studies have tended to focus only on the security of visual encoders or LLMs. On the visual encoder, \citet{wang2023instructta, yin2024vlattack} present adversaries with an opportunity to compromise model security by targeting solely the visual encoder. As for LLMs, previous studies \cite{carlini2023aligned,zou2023universal,chao2023jailbreaking} have fully explored the safety of LLM, and the user has the option of choosing a better performing and safer LLM on his or her own. However, the security of visual language projectors has not been fully explored, and only a few studies have focused on attacks on VLPs. Although some studies \cite{zhang2025qava} have attempted to attack the VLP (mainly Q-former), they have not thoroughly analyzed the security of the VLP structure. 

While compressed and uncompressed projectors each demonstrate distinct strengths in terms of computational efficiency and model performance, their respective trade-offs are particularly evident in large language and vision models. Compressed projectors have shown superior efficiency, particularly excelling in high-resolution visual understanding tasks and video comprehension scenarios. In contrast, uncompressed projectors maintain greater computational capabilities at the cost of higher computational expense. Previous studies \cite{yao2024deco} have extensively analyzed these trade-offs from perspectives of computational efficiency and model performance; however, a critical yet underexplored dimension in LVLM applications remains their security implications.
 
We assess the vulnerabilities of both compressed and uncompressed projectors by conducting adversarial attacks across diverse white-box and gray-box scenarios to evaluate the security implications for different VLP architectures. Our findings reveal three key insights: (1) Compressed projectors exacerbate security vulnerabilities beyond those affecting visual encoders. Specifically, an adversary can significantly degrade model performance by targeting compressed projectors, even with limited knowledge about the VLP. (2) In contrast, uncompressed projectors do not introduce additional security risks, as attacking them yields results comparable to attacking visual encoders directly. (3) This discrepancy in security between compressed and uncompressed projectors stems from their architectural differences and remains independent of the number of visual tokens. Notably, even when the visual tokens of an uncompressed projector are reduced (\eg, through pooling) to approach those of a compressed projector, the uncompressed projector still maintains its robust security characteristics.

Based on our analyses, we propose the following suggestions: (1) We strongly recommend employing uncompressed projectors rather than compressed projectors in high-security environments to mitigate potential risks from adversarial attacks. (2) In scenarios where computational efficiency is a priority, researchers can implement techniques like pooling operations to proportionally reduce the number of their output tokens. This approach offers enhanced security compared to using compressed projectors while maintaining acceptable performance levels, thereby striking an appropriate balance among effectiveness, efficiency, and security.

Our research provides three key contributions: (1) This study is \textbf{the first to systematically investigate VLP security}, presenting a comprehensive and systematic evaluation of various VLP architectures. (2) Our experimental results \textbf{uncover significant security vulnerabilities in compressed projectors}, revealing critical implications for models that employ such components. (3) We demonstrate that operating on visual tokens generated by uncompressed projectors offers enhanced security compared to compressed projectors when the number of visual tokens is largely comparable, thereby identifying promising approaches for VLP selection under efficiency constraints. 

\section{Related Work}
\label{sec:related_work}

\noindent
\textbf{Large Vision-Language Models.}
Training LVLMs from scratch is often prohibitively expensive. Thus, current popular frameworks typically begin with a pre-trained unimodal visual encoder and a large language model, focusing their training efforts on developing a VLP to connect the two modalities and effectively incorporate visual features into the LLM framework. The mainstream VLPs currently fall into two primary categories: compressed projectors and uncompressed projectors. Compressed projectors, such as Q-former \cite{li2023blip}, Resampler \cite{alayrac2022flamingo}, and D-Abstractor \cite{cha2024honeybee}, reduce the number of visual tokens to a fixed number of output tokens. In contrast, uncompressed projectors, exemplified by MLP \cite{liu2024visual}, maintain a proportional relationship between the number of visual tokens and their corresponding output tokens. 

Compressed and uncompressed projectors present distinct trade-offs in terms of computational efficiency, memory usage, and implementation complexity. On one hand, uncompressed projectors offer simplicity in design but incur significant computational costs. Specifically, Specifically, their computational requirements increase significantly as the token length scales linearly with the square of the resolution \cite{chen2024internvl}. Furthermore, in video processing applications \cite{ren2024timechat}, this scaling behavior results in a linear growth of required length relative to the number of video frames. On the other hand, compressed projectors achieve improved efficiency by reducing redundancy within the visual space. This is accomplished through the compression of token lengths from the visual encoder to a specified capacity, enabling strong performance while maintaining high computational efficiency. While prior research has extensively studied the performance and efficiency trade-offs between these two approaches in VLPs, there remains a critical gap: the safety implications of these methods have been largely neglected.

\noindent
\textbf{Adversarial Attacks on LVLMs.}
Prior research has shown that deep neural networks are vulnerable to adversarial perturbations \cite{szegedy2014intriguing, nguyen2015deep}. While extensive studies have investigated the security of VEs and LLMs \cite{shayegani2023survey, carlini2023aligned, liu2023jailbreaking, shayegani2023jailbreak}, relatively little attention has been devoted to examining the security of VLPs. Our work aims to address this gap by conducting adversarial attacks on both compressed and uncompressed projectors to evaluate their respective security properties.

This paper focuses on a practical gray-box scenario where the adversary is aware of both the VE weights and the structure of VLP, whether it be compressed or uncompressed. By crafting specialized loss functions tailored to the VLP architecture, the adversary can effectively execute targeted adversarial attacks against LVLMs in both white-box and gray-box settings. Our findings not only expose the vulnerabilities inherent in compressed projectors but also provide new insights into selecting between compressed and uncompressed projectors from a security perspective. 

\section{Method}
\label{sec:method}

\subsection{Preliminaries}
\noindent
\textbf{Notations}. We provide a brief overview of the definition of notations. An LVLM, denoted as $f$, generally comprises three key components: (1) the visual encoder $f_\text{VE}$, typically based on CLIP; (2) the vision-language projector $f_\text{VLP}$, which is usually implemented as a Q-former or an MLP; and (3) the large language model $f_\text{LLM}$. The LVLM $f$ accepts an image $x_i$ and an instruction $x_t$ as input, producing an output $y$. The process begins with the input image $x_i$ being processed by $f_\text{VE}$ to generate the visual feature representation $f_\text{VE}(x_i)$. Next, depending on whether it is a compressed or uncompressed VLP: (1) for the compressed VLP, $f_\text{VLP}$ extracts relevant features from both the visual features $f_\text{VE}(x_i)$ and the instruction $x_t$; while (2) for the uncompressed VLP, $f_\text{VLP}$ transforms the visual token dimension to align with the LLM input space. Both processes result in the projection output $f_\text{VLP}(f_\text{VE}(x_i), x_t)$. These projected features are then fed into $f_\text{LLM}$, which generates the final output $y$ as \cref{eq:LVLM}:
\begin{equation}
    \label{eq:LVLM}
    y = f(x_i, x_t) = f_\text{LLM}(f_\text{VLP}(f_\text{VE}(x_i), x_t), x_t).
\end{equation}

\noindent
\textbf{Compressed projectors}. The Q-former stands as the canonical representative of compressed projectors, functioning as a lightweight yet trainable querying transformer. It effectively extracts textually relevant features from the VE CLIP through a set of learnable queries. Notably, during BLIP-2's two-stage pre-training process, both the VE and the LLM remain static, with only the Q-former and its queries being trainable components. This training paradigm has been successfully employed in subsequent studies, such as InstructBLIP \cite{dai2305instructblip} and MiniGPT-4 \cite{zhu2023minigpt}, which also adopt similar approaches.

\noindent
\textbf{Uncompressed projectors}. A typical approach involves using a simple multilayer perceptron (MLP) to project the visual token representation into the input space of an LLM. For instance, in LLaVA v1.5 \cite{liu2024visual}, the visual encoder outputs a 1024-dimensional embedding, which is then mapped to 4096 dimensions via an MLP to align with the input requirements of the Vicuna LLM. Similarly, Eagle \cite{shi2024eagle} concatenates feature representations from multiple visual encoders to form a longer vector (typically ranging from 6000 to 8000 dimensions, depending on the number of visual encoders used), which is projected down to 4096 dimensions using an MLP.

\noindent
\textbf{The difference between compressed and uncompressed projectors}. We adopt the definition from \cite{yao2024deco}, where a compressed projector (represented by the Q-former) extracts information from visual tokens \emph{through a fixed number of query tokens}, with the output token count \emph{fixed to a relatively small number}. In contrast, the uncompressed projector, implemented as an MLP, processes and reshapes both the quantity and dimensionality of visual tokens. Let $N$ represent the number of tokens generated by a visual encoder. An uncompressed projector outputs tokens corresponding to fractions of $N$, specifically $N$, $N/2$, $N/4$, and so on. In contrast, a compressed projector generates a consistent number of tokens $M$ (typically where $M \ll N$).

\subsection{An Empirical Analysis of Component Accessibility Within LVLM Attacks}
\label{sec:accessibility_LVLM}

\noindent
\textbf{Visual encoder (VE)}. Current popular LVLMs, such as LLAVA \cite{liu2024visual}, BLIP-2 \cite{li2023blip}, InstructBLIP \cite{dai2305instructblip}, and MiniGPT-4 \cite{zhu2023minigpt}, employ either the ViT-L/14 model from CLIP \cite{radford2021learning} or the ViT-g/14 model from EVA-CLIP \cite{fang2023eva} as their visual encoders. Notably, none of these LVLMs fine-tune CLIP during training, which allows adversaries to easily access the visual encoder weights directly through publicly available CLIP checkpoints.

\noindent
\textbf{Vision-language Projector (VLP)}. We assume the adversary possesses knowledge of the VLP's architectural details, such as its implementation as a Q-former or MLP. Notably, all LVLMs built on Q-former, including BLIP-2, InstructBLIP, and MiniGPT-4, share identical VLP architectural designs. However, while the adversary may be aware of these structure specifics, obtaining the exact model parameters remains challenging due to the diversity in pre-trained LLMs and training datasets.

\noindent
\textbf{Large language model (LLM)}. The ecosystem of LLM is highly diverse, with many open-source options available, including OPT \cite{zhang2022opt}, LLaMA \cite{touvron2023llama}, FlanT5 \cite{chung2022scaling}, and Vicuna \cite{chiangvicuna}. Additionally, LVLMs often utilize customized datasets for parameter-efficient fine-tuning \cite{ding2023parameter} of these LLMs. Furthermore, numerous powerful closed-source LLMs are also accessible to model developers. This diversity in both open-source and closed-source LLMs presents a significant challenge for adversaries attempting to determine the specific architectural details or weights of the LLMs employed within LVLMs.

\noindent
\textbf{The white-box setting}. In the white-box setting, where adversaries have full access to model parameters (including both VEs and VLPs), we evaluate the safety of VLPs under worst-case scenarios. The adversary aims to modify a clean image to create an adversarial example capable of fooling the LVLM, leading to incorrect or unintended responses.

\noindent
\textbf{The gray-box setting}. Building on the analysis of LVLMs’ components (VE, VLP, and LLM) presented earlier, we focus on a practical yet hitherto unexplored gray-box setting, in which the adversary has access to the weights of the VE CLIP and the architectural structure of the VLP but lacks access to either the weights of the VLP or any information about the LLM. In this context, we can obtain a surrogate VLP model through transfer learning from other LVLMs, or alternatively, train the surrogate VLP model from scratch.

\subsection{Surrogate Models Used in Attacks}
Based on our analysis of the accessibility levels of various LVLM components presented in \cref{sec:accessibility_LVLM}, we design and implement adversarial attacks against VLPs under three distinct scenarios with increasing levels of difficulty to comprehensively assess their security vulnerabilities. (1) The first scenario represents the simplest case, where the attack operates in a white-box setting, allowing full visibility into the model's architecture and parameters. (2) The second scenario involves attacking a surrogate VLP from a similar LVLM, targeting another specific LVLM. (3) The most challenging scenario requires an attacker to develop a surrogate VLP from scratch, significantly increasing the complexity of generating effective adversarial examples. 
     
\noindent
\textbf{Surrogate models under white-box setting}. In our white-box setting, both the VE and VLP of the target model are accessible. This allows us to leverage $\mathcal{L}_\text{VE}$ and $\mathcal{L}_\text{VLP}$ to attack the target LVLMs.

\noindent
\textbf{Surrogate models transferred from other LVLMs in gray-box setting}. In our gray-box setting, while only the VE of the target model is available, we can still obtain surrogate VLPs by transferring knowledge from other similar LVLMs. For instance, if the target model is InstructBLIP Vicuna-13B, we can utilize the VLP from InstructBLIP $\text{FlanT5}_\text{XL}$ to attack it.

\noindent
\textbf{Surrogate models trained from scratch in gray-box setting}. When similar LVLMs are unavailable for transfer, the structure insights into VLP provide us with a unique opportunity. Once the target model's VLP structure (using compressed or uncompressed projectors) is identified, we can train surrogate Q-formers or MLPs from scratch. The detailed training procedure is outlined in \cref{sec:experiment_settings}. Notably, during this process, no information about the specific LLM used by the target LVLM is required; only the VE weights and VLP structure are necessary for training the surrogate VLP.

\begin{figure*}[t]
  \centering
  \includegraphics[width=0.92\textwidth]{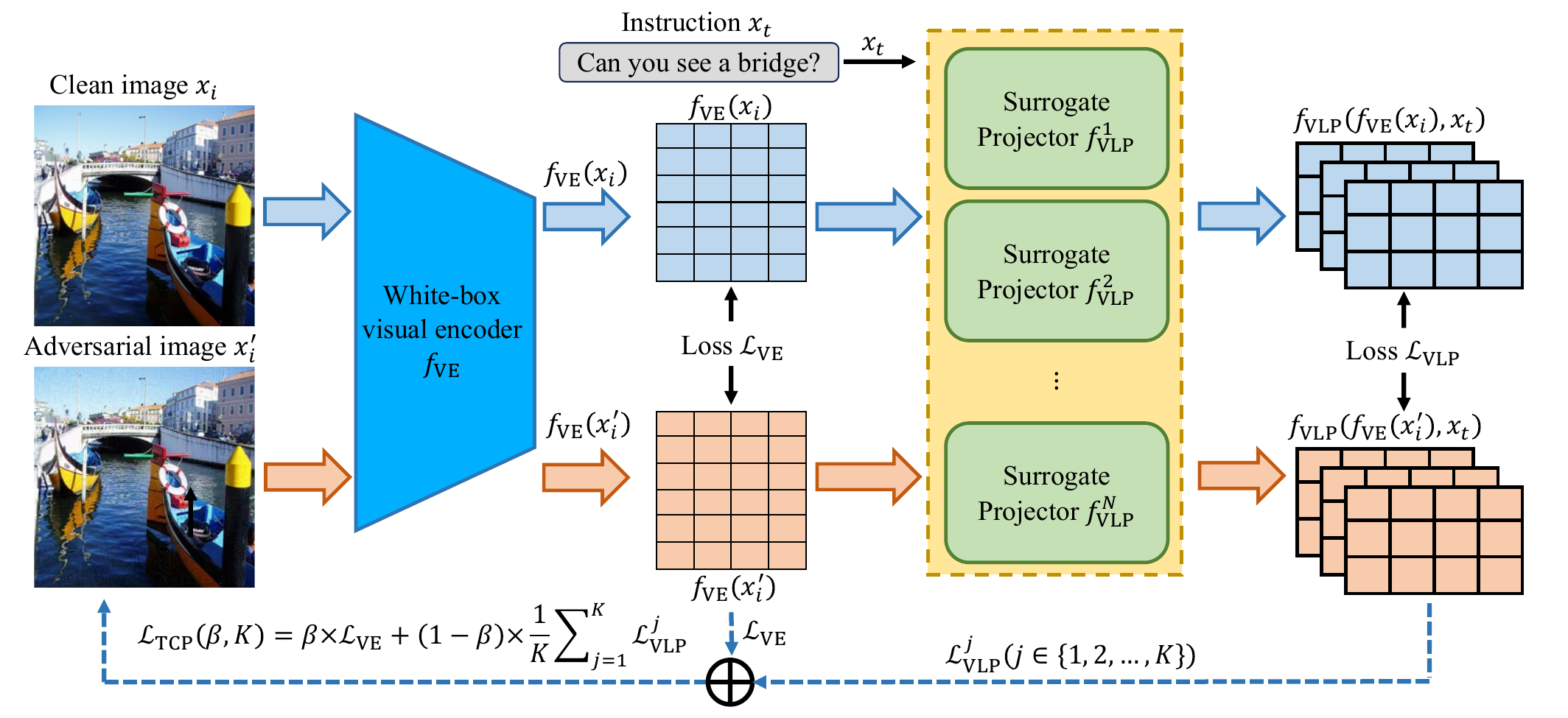}
  \caption{The attack pipeline operates by first extracting surrogate VLPs $\{f_\text{VLP}^1,\dots,f_\text{VLP}^N\}$, which are then utilized to generate adversarial examples $x'_i$ through the loss functions $\mathcal{L}_\text{VE}$, $\mathcal{L}_\text{VLP}$ and $\mathcal{L}_\text{TCP}(\beta, K)$ (TCP stands for ``Threat of Compressed Projectors''). A key aspect of this investigation is determining whether incorporating attacks on VLPs increases security vulnerabilities compared to solely attacking VEs, thereby assessing the robustness of the VLP structure.}
  \label{fig:TCP_overall}
\end{figure*}

\subsection{Loss Function for More Effective Attacks}

To evaluate the security of the VLP structure, we employ adversarial attack methods to analyze the robustness of LVLM. For a model with loss function $\mathcal{L}$, Fast Gradient Sign Method (FGSM) \cite{goodfellow2015explaining} performs an adversarial attack through gradient ascent as follows: $x'_i = x'_i + \nabla_{x'_i}\mathcal{L}$. In contrast, Projected Gradient Descent (PGD) \cite{madry2018towards} conducts multiple iterations of gradient ascent and projects $x'_i$ onto the constrained perturbation space after each step. Typically, PGD restricts the $l_\infty$ norm distance with $\|x_i - x'_i\|_\infty \leq \epsilon$. Additionally, a variant of the Carlini \& Wagner (C\&W) attack \cite{carlini2017towards}, referred to as CW-$l_2$, aims to maximize the loss function $\mathcal{L}$ while minimizing the $l_2$ norm distance $||x_i - x'_i||_2^2$.

To compare the security implications introduced by the VLP structure, we evaluate the attack results on VE and VLP. For a clean image input, let $V_\text{out} = f_\text{VE}(x_i)$ and $P_\text{out} = P(f_\text{VE}(x_i), x_t)$ denote the outputs of VE and VLP, respectively. For an adversarial image, let $V'_\text{out} = f_\text{VE}(x'_i)$ and $P'_\text{out} = P(f_\text{VE}(x'_i), x_t)$ represent the corresponding outputs, where $x'_i$ is initialized by adding random noise to $x_i$. We perform gradient-based adversarial attacks on both VE and VLP using the loss functions defined in \cref{eq:mse_v,eq:mse_q}, where $I$ and $J$ denote the sequence length and feature dimension:
\begin{align}
    \mathcal{L}_\text{VE} &= \frac{1}{IJ}\sum_I\sum_J(V_\text{out}-V'_\text{out})^2 \label{eq:mse_v} \\
    \mathcal{L}_\text{VLP} &= \frac{1}{IJ}\sum_I\sum_J(P_\text{out}-P'_\text{out})^2 \label{eq:mse_q}
\end{align}

The proposed loss function $\mathcal{L}_\text{TCP}$ (TCP stands for ``Threat of Compressed Projectors'') combines these components, with $\beta \in [0, 1]$ controlling the trade-off between VE and VLP losses, and $K$ denoting the number of surrogate VLPs used to enhance attack effectiveness. The overall loss function is:
\begin{equation}
    \mathcal{L}_\text{TCP} = \beta \mathcal{L}_\text{VE} + (1-\beta) \times \frac{1}{K} \Sigma_{j=1}^{K}\mathcal{L}_\text{VLP}^j
    \label{eq:tcp}
\end{equation}
In our experiments, we adopt $\beta = 0$ and $K=1$ by default, under which settings the loss functions $\mathcal{L}_\text{TCP}$ and $\mathcal{L}_\text{VLP}$ are mathematically equivalent. 

We evaluate the robustness of VLP by feeding adversarial samples generated from attacking both VE and VLP into it. If attacking VLP yields stronger adversarial effects, this indicates that integrating VLP increases LVLM's vulnerability compared to VE, suggesting that VLP is less secure in this context. Conversely, if attacking VLP proves less effective than attacking VE, this implies that VLP exhibits superior security performance.

\section{Experiments}
\label{sec:experiments}

\subsection{Experimental Settings}
\label{sec:experiment_settings}

\noindent
\textbf{Datasets}. We randomly sampled 2000 image-question pairs from VQA v2 \cite{antol2015vqa} to construct our primary dataset, referred to as ``VQA v2 2000''. Additionally, we include results for other datasets in \cref{sec:other_dataset_lvlm}, including ImageNet \cite{deng2009imagenet}, VizWiz \cite{bigham2010vizwiz}, and COCO \cite{lin2014microsoft}.

\noindent
\textbf{Evaluations}. For VQA v2, we used the official evaluation program to compute the VQA scores. For other tasks, we employed various metrics such as accuracy and CIDEr \cite{vedantam2015cider}.

\noindent
\textbf{Models}. For the compressed projector, we implemented four variants of InstructBLIP \cite{dai2305instructblip}, all of which employed the same ViT-g/14 vision encoder \cite{radford2021learning}. These models differed solely in their weight and bias parameters within the post-normalization layer. Notably, despite incorporating different LLMs, all these models maintained an identical VLP architecture. In contrast, our experiments with the uncompressed projector utilized a combined set of eight model variants from LLaVA-v1.5 \cite{liu2024visual} and LLaVA-v1.6 \cite{liu2024llavanext}. 

\noindent
\textbf{Adversarial attack}. We generated adversarial perturbations on clean images using PGD-$l_\infty$ \cite{madry2018towards} and CW-$l_2$ \cite{carlini2017towards}. For PGD-$l_\infty$, we set the attack step to 2/255, maximum perturbation to 8/255, and the number of attack steps to 20. For CW-$l_2$, we configured the attack step to 0.01 (equivalent to 2.55/255), with a constant of 0.005 and zero confidence.

\noindent
\textbf{Training details.}
(1) Surrogate compressed VLPs: we followed BLIP-2's two-stage pre-training process \cite{li2023blip}. In the first stage, vision-language representation learning was performed using a frozen VE. In the second stage, both the VE and LLM were kept frozen while performing vision-to-language generative learning. To ensure architectural diversity, we used different LLMs during pre-training (\eg, opt-2.7B and opt-6.7B) compared to the target LVLM. Pre-training was conducted using the COCO datasets.
% This stage included tasks such as Image-Text Matching (ITM), Image-Text Contrastive Learning (ITC), and Image-Grounded Text Generation (Image Captioning, IC). 
(2) Surrogate uncompressed VLPs: we followed LLAVA's official pre-training and fine-tuning settings. Similar to our compressed models, we used a different LLM during pre-training and fine-tuning to align with our gray-box experimental setup.

\noindent
\textbf{Multiple runs to ensure accuracy}. To mitigate the impact of random factors, each experiment was repeated five times using different random seeds. We report the mean $\mu$ of the results. Additionally, we measured the variance $\sigma$ across multiple runs, which generally remained low (around 0.1) with a maximum variance $\sigma_\text{max}\le0.25$.

\begin{table}[!htbp]
    \centering
    \resizebox{\linewidth}{!}{
    \begin{tabular}{c|c|c|c|c|c}
        \toprule
        Attack & \multirow{2}{*}{VLP} & \multirow{2}{*}{LVLM} & \multicolumn{3}{c}{VQA scores} \\
        method & & & Clean & $\mathcal{L}_\text{VE}$ & $\mathcal{L}_\text{VLP}$ \\
        \midrule
        \multirow{6}{*}{PGD} & \multirow{2}{*}{Uncompressed} & LLaVA-v1.5-7B & \cellcolor{gray!30}77.31 & 69.30 & 71.21 \\
        & & LLaVA-v1.5-13B & \cellcolor{gray!30}78.56 & 70.55 & 72.91 \\
        \cmidrule(lr){2-6}
        & \multirow{4}{*}{Compressed} & InstructBLIP-XL & \cellcolor{gray!30}72.43 & 43.51 & 34.15 \\
        & & InstructBLIP-7B & \cellcolor{gray!30}75.53 & 44.94 & 41.48 \\
        & & InstructBLIP-XXL & \cellcolor{gray!30}71.86 & 43.57 & 31.01 \\
        & & InstructBLIP-13B & \cellcolor{gray!30}68.22 & 42.02 & 36.96\\
        \midrule
        \multirow{6}{*}{CW} & \multirow{2}{*}{Uncompressed} & LLaVA-v1.5-7B & \cellcolor{gray!30}77.31 & 65.79 & 67.53 \\
        & & LLaVA-v1.5-13B & \cellcolor{gray!30}78.56 & 67.23 & 69.89 \\
        \cmidrule(lr){2-6}
        & \multirow{4}{*}{Compressed} & InstructBLIP-XL & \cellcolor{gray!30}72.43 & 41.02 & 15.85 \\
        & & InstructBLIP-7B & \cellcolor{gray!30}75.53 & 43.58 & 40.17 \\
        & & InstructBLIP-XXL & \cellcolor{gray!30}71.86 & 41.50 & 11.52 \\
        & & InstructBLIP-13B & \cellcolor{gray!30}68.22 & 39.95 & 33.95\\
        \bottomrule
    \end{tabular}
    }
    \caption{Results of attacks target VE and VLP under white-box setting. The difference in performance between $\mathcal{L}_\text{VE}$ and $\mathcal{L}_\text{VLP}$ highlights the security vulnerability of compressed projectors.}
    \label{tab:TCP_white}
\end{table}

\subsection{Results of White-box Setting}
\label{sec:exp_TCP_white_box}

The results of adversarial attacks under the white-box setting are summarized in \cref{tab:TCP_white}. Here, $\mathcal{L}_\text{VE}$ represents an adversarial attack specifically targeting the VE, as formally defined in \cref{eq:mse_v}, while $\mathcal{L}_\text{VLP}$ denotes an attack directed at the VLP, defined in \cref{eq:mse_q}. A key observation emerges under this setting: for compressed projectors, adversarial attacks targeting VLP achieve superior performance compared to those targeting VE. This discrepancy underscores the heightened security vulnerability of compressed projection. Conversely, in the case of uncompressed projectors, attacks against VLP yield results comparable to those targeting VE, demonstrating the enhanced robustness of uncompressed projectors. These initial findings from white-box attack experiments reveal critical security implications for compressed projectors.

\begin{table}[!htbp]
    \centering
    \resizebox{\linewidth}{!}{
    \begin{tabular}{c|c|c|c|c|c}
        \toprule
        Attack & The source of  & \multicolumn{4}{c}{Target models: LLaVA-v1.6 (uncompressed projector)} \\
        method & surrogate MLP & Vicuna-7B & Mistral-7b & Vicuna-13B & Hermes-Yi-34B \\
        \midrule
        \multicolumn{2}{c|}{Clean} & \cellcolor{gray!30}78.32 & \cellcolor{gray!30}79.10 & \cellcolor{gray!30}75.61 & \cellcolor{gray!30}79.07 \\
        \midrule
        \multirow{4}{*}{PGD} & $\mathcal{L}_\text{VE}$ & 70.55 & 72.31 & 68.92 & 71.27 \\
        \cmidrule{2-6}
        & LLaVA-v1.5-7B & 72.97 & 73.62 & 70.48 & 73.86 \\
        & LLaVA-v1.5-13B & 73.60 & 73.77 & 69.25 & 72.85 \\
        \midrule
        \multirow{4}{*}{CW} & $\mathcal{L}_\text{VE}$ & 67.72 & 67.22 & 62.56 & 67.37 \\
        \cmidrule{2-6}
        & LLaVA-v1.5-7B & 68.11 & 69.10 & 65.94 & 68.76 \\
        & LLaVA-v1.5-13B & 68.06 & 69.49 & 65.35 & 68.55 \\
        \midrule
        \midrule
        Attack & The source of  & \multicolumn{4}{c}{Target models: InstructBLIP (compressed projector)}\\
        method & surrogate Q-former & $\text{FlanT5}_\text{XL}$ & Vicuna-7B & $\text{FlanT5}_\text{XXL}$ & Vicuna-13B \\
        \midrule
        \multicolumn{2}{c|}{Clean} & \cellcolor{gray!30}72.43 & \cellcolor{gray!30}75.53 & \cellcolor{gray!30}71.86 & \cellcolor{gray!30}68.22 \\
        \midrule
        \multirow{3}{*}{PGD} & $\mathcal{L}_\text{VE}$ & 43.51 & 44.94 & 43.57 & 42.02 \\
        \cmidrule{2-6}
        & InstructBLIP-7B & 38.56 & 41.48 & 38.04 & 40.20 \\
        & InstructBLIP-13B & 38.89 & 39.76 & 39.16 & 36.96 \\
        \midrule
        \multirow{3}{*}{CW} & $\mathcal{L}_\text{VE}$ & 41.02 & 43.58 & 41.50 & 39.95 \\
        \cmidrule{2-6}
        & InstructBLIP-7B & 36.73 & 40.17 & 35.38 & 38.60 \\
        & InstructBLIP-13B & 37.18 & 38.93 & 35.81 & 33.95 \\
        \bottomrule
    \end{tabular}
    }
    \caption{Results of attacking uncompressed and compressed projectors using surrogate VLP models from other similar LVLMs. The results are divided into two sections: uncompressed (top) and compressed (bottom). The findings demonstrate that uncompressed projectors exhibit superior robustness compared to their compressed counterparts.}
    \label{tab:TCP_llm_mlp}
\end{table}

\subsection{Results of Attacks via Surrogate VLPs from Other LVLMs}
\label{sec:exp_TCP_transfer}

We employ the surrogate VLP from other similar LVLMs to launch attacks against our target models. Specifically, for compressed projectors, we conduct experiments using InstructBLIP models that have been trained with various LLMs. For uncompressed projectors, we utilize LLaVA v1.5 as a surrogate model to attack LLaVA v1.6. As demonstrated in \cref{tab:TCP_llm_mlp}, our results indicate that uncompressed projectors maintain superior robustness under gray-box attack conditions, whereas attacks targeting compressed projectors lead to further degradation in model performance.

\subsection{Results of Attacks via Surrogate VLPs Trained from Scratch}
\label{sec:exp_TCP_blip2}

We further explored training the surrogate VLP from scratch, using a different LLM than the target LVLM within the context of our gray-box attack scenario. Our experimental results, as evidenced by \cref{tab:TCP_blip2}, demonstrate consistent findings with those presented in \cref{tab:TCP_llm_mlp}: uncompressed projectors exhibit superior security performance even under more challenging attacking scenarios.

\begin{table}[!htbp]
    \centering
    \resizebox{\linewidth}{!}{
    \begin{tabular}{c|c|c|c|c|c}
        \toprule
        Attack & Surrogate & \multicolumn{4}{c}{Target models: LLaVA-v1.6 (uncompressed projector)} \\
        method & MLP & Vicuna-7B & Mistral-7b & Vicuna-13B & Hermes-Yi-34B \\
        \midrule
        \multicolumn{2}{c|}{Clean} & \cellcolor{gray!30}78.32 & \cellcolor{gray!30}79.10 & \cellcolor{gray!30}75.61 & \cellcolor{gray!30}79.07 \\
        \midrule
        \multirow{4}{*}{PGD} & $\mathcal{L}_\text{VE}$ & 70.55 & 72.31 & 68.92 & 71.27 \\
        \cmidrule{2-6}
        & Vicuna-v1.3-7B & 72.55 & 73.06 & 69.60 & 73.58 \\
        & Vicuna-v1.3-13B & 73.71 & 73.76 & 69.66 & 73.24 \\
        \midrule
        \multirow{4}{*}{CW} & $\mathcal{L}_\text{VE}$ & 67.72 & 67.22 & 62.56 & 67.37 \\
        \cmidrule{2-6}
        & Vicuna-v1.3-7B & 69.28 & 71.03 & 66.09 & 70.18 \\
        & Vicuna-v1.3-13B & 69.53 & 70.12 & 66.67 & 69.49 \\
        \midrule
        \midrule
        Attack & Surrogate & \multicolumn{4}{c}{Target models: InstructBLIP (compressed projector)} \\
        method &  Q-former & $\text{FlanT5}_\text{XL}$ & Vicuna-7B & $\text{FlanT5}_\text{XXL}$ & Vicuna-13B \\
        \midrule
        \multicolumn{2}{c|}{Clean} & \cellcolor{gray!30}72.43 & \cellcolor{gray!30}75.53 & \cellcolor{gray!30}71.86 & \cellcolor{gray!30}68.22 \\
        \midrule
        \multirow{3.5}{*}{PGD} & $\mathcal{L}_\text{VE}$ & 43.51 & 44.94 & 43.57 & 42.02 \\
        \cmidrule{2-6}
        \cmidrule{2-6}
        \cmidrule{2-6}
        & opt-2.7B & 39.82 & 41.81 & 39.89 & 39.66 \\
        \cmidrule{2-6}
        & opt-6.7B & 40.06 & 41.79 & 40.32 & 39.31 \\
        \midrule
        \multirow{3.5}{*}{CW} & $\mathcal{L}_\text{VE}$ & 41.02 & 43.58 & 41.50 & 39.95\\
        \cmidrule{2-6}
        \cmidrule{2-6}
        \cmidrule{2-6}
        & opt-2.7B & 40.39 & 41.96 & 39.53 & 39.61 \\
        \cmidrule{2-6}
        & opt-6.7B & 39.82 & 41.81 & 39.89 & 39.66 \\
        \bottomrule
    \end{tabular}
    }
    \caption{Results on uncompressed/compressed projectors using a surrogate VLP model trained from scratch.}
    \label{tab:TCP_blip2}
\end{table}

\subsection{Is Security Dependent on Architecture or the Quantity of Visual Tokens?}
\label{sec:exp_tokens}

Our experimental results, as summarized in \cref{tab:TCP_white,tab:TCP_llm_mlp,tab:TCP_blip2}, demonstrate that \textbf{\emph{compressed projectors exhibit significant security vulnerabilities compared to their uncompressed counterparts}}. However, it is important to note that uncompressed projectors generally possess a larger number of tokens (\eg, the MLP employed in our laboratory utilizes 576 visual tokens, whereas the Q-former operates with only 32 visual tokens). This raises an essential question: does the reduced security of the Q-former stem from the inherent characteristics within its architecture, or merely from its smaller quantity of visual tokens? 

To investigate this, we systematically apply mean pooling operations to LLaVA's original 576 (24x24) visual tokens. First, a 2x2 mean pooling operation is performed, reducing the number of visual tokens to 144. Subsequently, we further apply a 4x4 mean pooling operation, compressing the visual tokens down to 36.  

We then conduct adversarial attacks on LLaVA-v1.5-7B with varying numbers of visual tokens. The results in \cref{tab:TCP_reduce_token_llava} reveal that even when only 36 visual tokens are utilized (comparable to Q-former's), the attack performance on VLP does not surpass that on VE, implying its robustness against attacks. This finding conclusively demonstrates that the enhanced robustness of uncompressed projectors is \textbf{\emph{not merely dependent on the quantity of visual tokens, but rather arises inherently from their structure design}}. Furthermore, this implies that in pursuit of efficiency, one can opt for an uncompressed projector while employing techniques such as spatial pooling to minimize its visual markers. Even when their respective quantities of visual tokens are comparable in number, the security provided by an uncompressed projector remains markedly superior to that of its compressed ones.

\begin{table}[!htbp]
    \centering
    \resizebox{0.84\linewidth}{!}{
    \begin{tabular}{c|c|c|c|c}
        \toprule
        Attack & LVLM & \multicolumn{3}{c}{VQA scores} \\
        method & visual token number & Clean & $\mathcal{L}_\text{VE}$ & $\mathcal{L}_\text{VLP}$ \\
        \midrule
        \multirow{3}{*}{PGD} & 576 (official) & \cellcolor{gray!30}{77.31} & 69.30 & 71.21 \\
        & 144 & \cellcolor{gray!30}{76.03} & 68.63 & 69.19 \\
        & 36 & \cellcolor{gray!30}{73.09} & 66.72 & 67.30 \\
        \midrule
        \multirow{3}{*}{CW} & 576 (official) & \cellcolor{gray!30}{77.31} & 65.79 & 67.53 \\
        & 144 & \cellcolor{gray!30}{76.03} & 63.06 & 65.25 \\
        & 36 & \cellcolor{gray!30}{73.09} & 61.97 & 64.12 \\
        \bottomrule
    \end{tabular}
    }
    \caption{Results of the attack on LLaVA with largely reduced visual tokens under white-box settings.}
    \label{tab:TCP_reduce_token_llava}
\end{table}

\subsection{Further Strategies to Deteriorate the Security of Compressed Projectors}
\label{sec:exp_TCP_multi}

\Cref{tab:TCP_blip2} presents the attack via training surrogate VLPs. We employ a single surrogate VLP for simplicity. However, multiple training runs can yield additional surrogate models, which we then leverage collectively to launch attacks against the target model, as illustrated in \cref{fig:TCP_overall}. The results summarized in \cref{tab:TCP_blip2_num} demonstrate that this multi-VLPs attack strategy further degrades the performance of the compressed projector.

Notably, increasing the number of surrogate VLPs does not significantly escalate the computational complexity of the attack. This is because all surrogate VLPs share a common visual encoder, and each Q-former component is parameter-efficient, containing approximately one-fourth the parameters of the VE. Consequently, for $N=2$ surrogates, the total parameter count increases to roughly 1.2 times that of $N=1$, while for $N=3$, it reaches approximately 1.4 times that of $N=1$.

\begin{table}[!htbp]
    \centering
    \resizebox{\linewidth}{!}{
    \begin{tabular}{c|c|c|c|c|c|c}
        \toprule
        Attack & \multirow{2}{*}{$K$} & \multicolumn{4}{c|}{The target models (InstructBLIP)} & \multirow{2}{*}{Avg.} \\
        method & & $\text{FlanT5}_\text{XL}$ & Vicuna-7B & $\text{FlanT5}_\text{XXL}$ & Vicuna-13B & \\
        \midrule
        \multicolumn{2}{c|}{baseline} & \cellcolor{gray!30}72.43 & \cellcolor{gray!30}75.53 & \cellcolor{gray!30}71.86 & \cellcolor{gray!30}68.22 & \cellcolor{gray!30}71.51 \\
        \midrule
        \multirow{4}{*}{PGD} & $\mathcal{L}_\text{VE}$ & 43.51 & 44.94 & 43.57 & 42.02 & 43.51 \\
        \cmidrule{2-7}
        & 1 & 40.06 & 41.79 & 40.32 & 39.31 & 40.37 \\
        & 2 & 39.45 & 40.57 & 39.15 & 38.64 & 39.45 \\
        & 3 & \cellcolor{yellow!60}37.58 & \cellcolor{yellow!60}39.80 & \cellcolor{yellow!60}38.07 & \cellcolor{yellow!60}37.32 & \cellcolor{yellow!60}38.19 \\
        \midrule
        \multirow{4}{*}{CW} & $\mathcal{L}_\text{VE}$ & 41.02 & 43.58 & 41.50 & 39.95 & 41.51 \\
        \cmidrule{2-7}
        & 1 & 39.82 & 41.81 & 39.89 & 39.66 & 40.30 \\
        & 2 & 38.08 & 39.41 & 38.49 & 37.66 & 38.41 \\
        & 3 & \cellcolor{yellow!60}37.13 & \cellcolor{yellow!60}38.99 & \cellcolor{yellow!60}37.49 & \cellcolor{yellow!60}36.58 & \cellcolor{yellow!60}37.55 \\
        \bottomrule
    \end{tabular}
    }
    \caption{Results of attacks using multi-VLPs.}
    \label{tab:TCP_blip2_num}
\end{table}

We start by setting $\beta=0$ to specifically examine attacks designed to target VLP, enabling us to analyze whether incorporating VLP improves attack effectiveness. We then combine the $\mathcal{L}_\text{VE}$ and $\mathcal{L}_\text{VLP}$ losses as \cref{eq:tcp}. As shown in \cref{tab:TCP_blip2_vision}, $\mathcal{L}_\text{TCP}$ achieves superior attack performance compared to other approaches. Additionally, some adversarial images are provided in \cref{sec:demos}.

\begin{table}[!htbp]
    \centering
    \resizebox{\linewidth}{!}{
    \begin{tabular}{c|c|c|c|c|c|c|c}
        \toprule
        Attack & \multirow{2}{*}{$K$} & \multirow{2}{*}{$\beta$} & \multicolumn{4}{c|}{The target models (InstructBLIP)} & \multirow{2}{*}{Avg.} \\
        method & & & $\text{FlanT5}_\text{XL}$ & Vicuna-7B & $\text{FlanT5}_\text{XXL}$ & Vicuna-13B & \\
        \midrule
        \multicolumn{3}{c|}{baseline} & \cellcolor{gray!30}72.43 & \cellcolor{gray!30}75.53 & \cellcolor{gray!30}71.86 & \cellcolor{gray!30}68.22 & \cellcolor{gray!30}71.51 \\
        \midrule
        \multirow{6}{*}{PGD} & $\mathcal{L}_\text{VE}$ & 1 & 43.51 & 44.94 & 43.57 & 42.02 & 43.51 \\
        \cmidrule{2-8}
        & \multirow{5}{*}{2} & 0.4 & 37.87 & 38.89 & 38.68 & 37.85 & 38.32 \\
        & & 0.3 & 38.29 & 40.08 & 38.38 & 37.79 & 38.64 \\
        & & 0.2 & \cellcolor{yellow!60}36.91 & \cellcolor{yellow!60}39.71 & \cellcolor{yellow!60}37.46 & \cellcolor{yellow!60}36.70 & \cellcolor{yellow!60}37.70 \\
        & & 0.1 & 37.64 & 39.82 & 37.72 & 38.38 & 38.39 \\
        & & 0 & 39.45 & 40.57 & 39.15 & 38.64 & 39.45 \\
        \midrule
        \multirow{6}{*}{CW} & $\mathcal{L}_\text{VE}$ & 1 & 41.02 & 43.58 & 41.50 & 39.95 & 41.51 \\
        \cmidrule{2-8}
        & \multirow{5}{*}{2} & 0.4 & 37.22 & 37.91 & 36.80 & 36.29 & 37.06 \\
        & & 0.3 & \cellcolor{yellow!60}35.78 & \cellcolor{yellow!60}38.18 & \cellcolor{yellow!60}36.87 & \cellcolor{yellow!60}36.88 & \cellcolor{yellow!60}36.93 \\
        & & 0.2 & 36.85 & 38.64 & 36.32 & 36.82 & 37.16 \\
        & & 0.1 & 36.54 & 39.17 & 37.33 & 37.32 & 37.59 \\
        & & 0 & 38.08 & 39.41 & 38.49 & 37.66 & 38.41 \\
        \bottomrule
    \end{tabular}
    }
    \caption{Results of attacks combined the $\mathcal{L}_\text{VE}$ and $\mathcal{L}_\text{VLP}$.}
    \label{tab:TCP_blip2_vision}
\end{table}

\subsection{Results on Additional Datasets}

\label{sec:other_dataset_lvlm}

Using InstructBLIP-based Vicuna-7B, we performed a thorough security evaluation to analyze potential vulnerabilities of compressed projectors across diverse datasets extending beyond VQA v2 in \cref{tab:TCP_blip2_more_dataset}. Our assessment revealed that in a gray-box attack scenario, where only the target model's VE weights and VLP architecture are known, we were able to significantly degrade model performance, highlighting the inherent security risks associated with compressed projectors.

\begin{table}[!htbp]
    \centering
    \resizebox{0.72\linewidth}{!}{
    \begin{tabular}{c|c|c}
        \toprule
        Dataset & Clean & Adv. \\
        \midrule
        ImageNet-1K accuracy & 81.0 & 14.9 \\
        VizWiz test-dev scores & 33.08 & 10.38 \\
        COCO CIDEr & 140.6 & 11.1 \\ 
        \bottomrule
    \end{tabular}
    }
    \caption{Attack InstructBLIP Vicuna-7B with surrogate VLP trained using opt-6.7B on other datasets and tasks.}
    \label{tab:TCP_blip2_more_dataset}
\end{table}

\noindent

\subsection{Ablation Study on Pre-Training Tasks}
\label{sec:exp_blip2_tasks}
BLIP-2's first-stage pre-training consists of three tasks: Image-Text Matching (ITM), Image-Text Contrastive Learning (ITC), and Image-Grounded Text Generation (Image Captioning, IC). By default, we adopted the standard BLIP-2 training configuration, employing all three tasks concurrently. However, our experimental results presented in \cref{tab:TCP_blip2_task} reveal an important insight: training the surrogate VLP exclusively with the IC task achieves attack performance comparable to utilizing all three tasks together. This not only reduces the overhead associated with training the surrogate model but also amplifies the security risks posed by the compressed projector.

\begin{table}[!htbp]
    \centering
    \resizebox{\linewidth}{!}{
    \begin{tabular}{c|c|c|c|c|c|c|c|c}
        \toprule
        Attack & \multirow{2}{*}{ITC} & \multirow{2}{*}{ITM} & \multirow{2}{*}{IC} & \multicolumn{4}{c|}{The target models (InstructBLIP)} & \multirow{2}{*}{Avg.}\\
        method & & & & $\text{FlanT5}_\text{XL}$ & Vicuna-7B & $\text{FlanT5}_\text{XXL}$ & Vicuna-13B & \\
        \midrule
        \multicolumn{4}{c|}{baseline} & \cellcolor{gray!30}72.43 & \cellcolor{gray!30}75.53 & \cellcolor{gray!30}71.86 & \cellcolor{gray!30}68.22 & \cellcolor{gray!30}71.51\\
        \midrule
        \multirow{7}{*}{PGD} & \ding{51} &  &  & 43.93 & 45.64 & 42.81 & 42.75 & 43.78 \\
        &  & \ding{51} &  & 48.84 & 49.66 & 47.88 & 46.31 & 48.17 \\
        &  &  & \ding{51} & 41.04 & 41.80 & 40.81 & 39.16 & 40.70 \\
        & \ding{51} & \ding{51} &  & 40.46 & 42.74 & 41.20 & 40.70 & 41.28 \\
        & \ding{51} &  & \ding{51} & 39.55 & 42.18 & 40.26 & 39.90 & 40.47 \\
        &  & \ding{51} & \ding{51} & 41.32 & 42.49 & 40.57 & 39.65 & 41.01 \\
        \cmidrule{2-9}
        & \ding{51} & \ding{51} & \ding{51} & 40.06 & 41.79 & 40.32 & 39.31 & 40.37 \\
        \midrule
        \multirow{7}{*}{CW} & \ding{51} &  &  & 41.55 & 44.11 & 40.75 & 41.32 & 41.93 \\
        &  & \ding{51} &  & 45.17 & 47.21 & 45.49 & 44.62 & 45.62 \\
        &  &  & \ding{51} & 40.64 & 41.81 & 39.78 & 38.63 & 40.22 \\
        & \ding{51} & \ding{51} &  & 39.52 & 42.58 & 39.88 & 39.32 & 40.33 \\
        & \ding{51} &  & \ding{51} & 39.47 & 41.84 & 40.28 & 38.74 & 40.08 \\
        &  & \ding{51} & \ding{51} & 40.17 & 41.99 & 40.10 & 40.23 & 40.62 \\
        \cmidrule{2-9}
        & \ding{51} & \ding{51} & \ding{51} & 39.82 & 41.81 & 39.89 & 39.66 & 40.30 \\
        \bottomrule
    \end{tabular}
    }
    \caption{Attack performance of the surrogate VLP with various pre-training task combinations. It suffices to employ IC task alone to achieve strong attacks.}
    \label{tab:TCP_blip2_task}
\end{table}

\subsection{Results on Various Attack Methods}

In addition to PGD and CW in \cref{tab:TCP_blip2}, we also implemented an additional attack method: MI-FGSM \cite{dong2018boosting}. Under gray-box settings, we conducted MI-FGSM attacks and extended the analysis presented in \cref{tab:TCP_blip2} of the original paper to incorporate methods beyond PGD and CW. The specific parameter configuration for MI-FGSM included setting the momentum value to 0.9. Our findings remain consistent with the conclusions in \cref{sec:exp_TCP_transfer,sec:exp_TCP_blip2}. Specifically, uncompressed projectors demonstrate superior robustness under gray-box attack settings, while attacks targeting compressed projectors result in further degradation of model performance.

\begin{table}[!htbp]
    \centering
    \resizebox{\linewidth}{!}{
    \begin{tabular}{c|c|c|c|c|c}
        \toprule
        Attack & Surrogate & \multicolumn{4}{c}{Target models: LLaVA-v1.6 (uncompressed projector)} \\
        method & MLP & Vicuna-7B & Mistral-7b & Vicuna-13B & Hermes-Yi-34B \\
        \midrule
        \multicolumn{2}{c|}{Clean} & \cellcolor{gray!30}78.32 & \cellcolor{gray!30}79.10 & \cellcolor{gray!30}75.61 & \cellcolor{gray!30}79.07 \\
        \midrule
        \multirow{4}{*}{MI-FGSM} & $\mathcal{L}_\text{VE}$ & 71.65 & 68.66 & 72.85 & 72.01 \\
        \cmidrule{2-6}
        & Vicuna-v1.3-7B & 72.40 & 69.24 & 73.48 & 73.10 \\
        & Vicuna-v1.3-13B & 72.90 & 69.99 & 73.58 & 73.49 \\
        \midrule
        \midrule
        Attack & Surrogate & \multicolumn{4}{c}{Target models: InstructBLIP (compressed projector)} \\
        method &  Q-former & $\text{FlanT5}_\text{XL}$ & Vicuna-7B & $\text{FlanT5}_\text{XXL}$ & Vicuna-13B \\
        \midrule
        \multicolumn{2}{c|}{Clean} & \cellcolor{gray!30}72.43 & \cellcolor{gray!30}75.53 & \cellcolor{gray!30}71.86 & \cellcolor{gray!30}68.22 \\
        \midrule
        \multirow{3.5}{*}{MI-FGSM} & $\mathcal{L}_\text{VE}$ & 49.99 & 52.41 & 49.40 & 47.06 \\
        \cmidrule{2-6}
        & opt-6.7B & 41.71 & 44.95 & 41.95 & 41.54 \\
        \bottomrule
    \end{tabular}
    }
    \caption{Additional results of \cref{tab:TCP_blip2} on MI-FGSM.} 
    \label{tab:TCP_blip2_MI_fgsm}
\end{table}

\section{Conclusion}
\label{sec:conclusion}
This study investigates the security vulnerabilities of VLP structures within two representative LVLM frameworks. Our analysis exposes severe security susceptibilities in compressed projectors, while highlighting the robust performance of uncompressed alternatives. Through rigorous empirical evaluation, we demonstrate that this vulnerability originates inherently from the architectural design of the compressed projector itself and remains independent of the visual token quantity. These findings sound a cautionary note regarding the security implications for LVLMs employing compressed projectors, while encouraging researchers to adopt a more comprehensive understanding of VLP performance characteristics.

\section{Discussion of Defense Mechanism}

We briefly discuss two potential strategies for enhancing model robustness:

\noindent
(1) \textbf{Optimizing uncompressed projector designs for improved efficiency}. While uncompressed projectors have demonstrated strong security properties, their inefficiency stems from a large number of visual tokens. However, our experiments in \cref{tab:TCP_reduce_token_llava} reveal that not all visual tokens are essential for performance. By applying 2x2 pooling to LLaVA vision tokens, we successfully reduced the number of visual tokens to 25\% of the original amount, resulting in only a marginal 1.3\% drop in accuracy, and the LVLMs are still robust. This suggests significant potential for optimizing uncompressed projectors by eliminating redundant visual tokens while retaining their inherent security benefits, thereby enhancing the robustness.

\noindent
(2) \textbf{Hybrid architectures combining compressed and uncompressed projectors}. Compressed projectors offer high efficiency and sufficient accuracy but lack robustness, whereas uncompressed projectors provide strong robustness and accuracy at the cost of efficiency. Building on insights from \cref{tab:TCP_reduce_token_llava}, pooling significantly reduced visual tokens, and the accuracy does not significantly decrease while maintaining safety, we propose a hybrid approach. For instance, we could pool the visual tokens from an uncompressed projector (originally 576 tokens) to 25\% of their original count (144 tokens), and supplement this with 32 tokens from a compressed projector, resulting in a total of 144+32=176 tokens. This combination aims to achieve a balance of high accuracy, efficiency, and robustness. Unfortunately, due to challenges in training LVLMs with compressed projectors (lack of fully open-sourced training code of Q-former based LVLMs), we may face limitations in fully implementing this.

\section{Limitations}
We summarize the limitations of our study as follows: (1) The attack methodology has certain constraints. We implemented the attacks using only basic PGD, CW and MI-FGSM methods on VQA and image captioning tasks. Notably, we did not employ advanced techniques designed to enhance adversarial transferability, \eg, DI-FGSM \cite{xie2019improving}. Additionally, we limited our attacks to these two tasks without extending to other domains. Despite this, we successfully exposed the security vulnerabilities of compressed projectors, and incorporating DI and MI methods could potentially amplify the robustness risks associated with these models. (2) Our study focuses exclusively on analyzing the security vulnerabilities of both compressed and uncompressed projectors. However, we do not investigate potential defense mechanisms against the attacks. A discussion of possible defensive strategies would provide valuable insights into enhancing the robustness of compressed projectors against such attacks.

\section{Acknowledgments}
This work was supported by Young Elite Scientists Sponsorship Program by CAST (2023QNRC001), the National Key R\&D Program of China (2023YFB4502200), the National Natural Science Foundation of China (No. 62376024, 62325405, 62104128, 62203257, 62031017, 62406159, U21B2031), Tsinghua University Initiative Scientific Research Program,  Beijing National Research Center for Information Science, Technology (No. BNR2024TD03001) and Beijing Innovation Center for Future Chips.

% Bibliography entries for the entire Anthology, followed by custom entries
%\bibliography{anthology,custom}
% Custom bibliography entries only
\clearpage
\bibliography{custom}

\begin{thebibliography}{41}
\providecommand{\natexlab}[1]{#1}

\bibitem[{Alayrac et~al.(2022)Alayrac, Donahue, Luc et~al.}]{alayrac2022flamingo}
Jean-Baptiste Alayrac, Jeff Donahue, Pauline Luc, and 1 others. 2022.
\newblock \href {https://openreview.net/forum?id=EbMuimAbPbs} {Flamingo: a visual language model for few-shot learning}.
\newblock In \emph{Advances in Neural Information Processing Systems}.

\bibitem[{Antol et~al.(2015)Antol, Agrawal, Lu, Mitchell, Batra, Zitnick, and Parikh}]{antol2015vqa}
Stanislaw Antol, Aishwarya Agrawal, Jiasen Lu, Margaret Mitchell, Dhruv Batra, C~Lawrence Zitnick, and Devi Parikh. 2015.
\newblock Vqa: Visual question answering.
\newblock In \emph{Proceedings of the IEEE international conference on computer vision}, pages 2425--2433.

\bibitem[{Bai et~al.(2025)Bai, Chen, Liu, Wang, Ge, Song, Dang, Wang, Wang, Tang et~al.}]{bai2025qwen2}
Shuai Bai, Keqin Chen, Xuejing Liu, Jialin Wang, Wenbin Ge, Sibo Song, Kai Dang, Peng Wang, Shijie Wang, Jun Tang, and 1 others. 2025.
\newblock Qwen2. 5-vl technical report.
\newblock \emph{arXiv preprint arXiv:2502.13923}.

\bibitem[{Bigham et~al.(2010)Bigham, Jayant, Ji, Little, Miller, Miller, Miller, Tatarowicz, White, White et~al.}]{bigham2010vizwiz}
Jeffrey~P Bigham, Chandrika Jayant, Hanjie Ji, Greg Little, Andrew Miller, Robert~C Miller, Robin Miller, Aubrey Tatarowicz, Brandyn White, Samual White, and 1 others. 2010.
\newblock Vizwiz: nearly real-time answers to visual questions.
\newblock In \emph{Proceedings of the 23nd annual ACM symposium on User interface software and technology}, pages 333--342.

\bibitem[{Carlini et~al.(2023)Carlini, Nasr, Choquette-Choo, Jagielski, Gao, Koh, Ippolito, Tram{\`e}r, and Schmidt}]{carlini2023aligned}
Nicholas Carlini, Milad Nasr, Christopher~A Choquette-Choo, Matthew Jagielski, Irena Gao, Pang~Wei Koh, Daphne Ippolito, Florian Tram{\`e}r, and Ludwig Schmidt. 2023.
\newblock Are aligned neural networks adversarially aligned?
\newblock In \emph{Advances in Neural Information Processing System (NeurIPS)}.

\bibitem[{Carlini and Wagner(2017)}]{carlini2017towards}
Nicholas Carlini and David Wagner. 2017.
\newblock \href {https://doi.org/10.1109/SP.2017.49} {{ Towards Evaluating the Robustness of Neural Networks }}.
\newblock In \emph{2017 IEEE Symposium on Security and Privacy (SP)}, pages 39--57.

\bibitem[{Cha et~al.(2024)Cha, Kang, Mun, and Roh}]{cha2024honeybee}
Junbum Cha, Wooyoung Kang, Jonghwan Mun, and Byungseok Roh. 2024.
\newblock Honeybee: Locality-enhanced projector for multimodal llm.
\newblock In \emph{Proceedings of the IEEE/CVF Conference on Computer Vision and Pattern Recognition}, pages 13817--13827.

\bibitem[{Chao et~al.(2023)Chao, Robey, Dobriban, Hassani, Pappas, and Wong}]{chao2023jailbreaking}
Patrick Chao, Alexander Robey, Edgar Dobriban, Hamed Hassani, George~J Pappas, and Eric Wong. 2023.
\newblock Jailbreaking black box large language models in twenty queries.
\newblock \emph{arXiv preprint arXiv:2310.08419}.

\bibitem[{Chen et~al.(2024)Chen, Wu, Wang, Su, Chen, Xing, Zhong, Zhang, Zhu, Lu et~al.}]{chen2024internvl}
Zhe Chen, Jiannan Wu, Wenhai Wang, Weijie Su, Guo Chen, Sen Xing, Muyan Zhong, Qinglong Zhang, Xizhou Zhu, Lewei Lu, and 1 others. 2024.
\newblock Internvl: Scaling up vision foundation models and aligning for generic visual-linguistic tasks.
\newblock In \emph{Proceedings of the IEEE/CVF conference on computer vision and pattern recognition}, pages 24185--24198.

\bibitem[{Chiang et~al.(2023)Chiang, Li, Lin, Sheng, Wu, Zhang, Zheng, Zhuang, Zhuang, Gonzalez et~al.}]{chiangvicuna}
Wei-Lin Chiang, Zhuohan Li, Zi~Lin, Ying Sheng, Zhanghao Wu, Hao Zhang, Lianmin Zheng, Siyuan Zhuang, Yonghao Zhuang, Joseph~E Gonzalez, and 1 others. 2023.
\newblock Vicuna: An open-source chatbot impressing gpt-4 with 90\%* chatgpt quality.
\newblock \emph{See https://vicuna.lmsys.org}.

\bibitem[{Chung et~al.(2024)Chung, Hou, Longpre, Zoph, Tay, Fedus, Li, Wang, Dehghani, Brahma et~al.}]{chung2022scaling}
Hyung~Won Chung, Le~Hou, Shayne Longpre, Barret Zoph, Yi~Tay, William Fedus, Yunxuan Li, Xuezhi Wang, Mostafa Dehghani, Siddhartha Brahma, and 1 others. 2024.
\newblock Scaling instruction-finetuned language models.
\newblock \emph{Journal of Machine Learning Research}, 25(70):1--53.

\bibitem[{Dai et~al.(2023)Dai, Li, Li, Tiong, Zhao, Wang, Li, Fung, and Hoi}]{dai2305instructblip}
W~Dai, J~Li, D~Li, AMH Tiong, J~Zhao, W~Wang, B~Li, P~Fung, and S~Hoi. 2023.
\newblock Instructblip: Towards general-purpose vision-language models with instruction tuning.
\newblock \emph{arXiv preprint arXiv:2305.06500}.

\bibitem[{Deng et~al.(2009)Deng, Dong, Socher, Li, Li, and Fei-Fei}]{deng2009imagenet}
Jia Deng, Wei Dong, Richard Socher, Li-Jia Li, Kai Li, and Li~Fei-Fei. 2009.
\newblock Imagenet: A large-scale hierarchical image database.
\newblock In \emph{2009 IEEE conference on computer vision and pattern recognition}, pages 248--255. Ieee.

\bibitem[{Ding et~al.(2023)Ding, Qin, Yang, Wei, Yang, Su, Hu, Chen, Chan, Chen et~al.}]{ding2023parameter}
Ning Ding, Yujia Qin, Guang Yang, Fuchao Wei, Zonghan Yang, Yusheng Su, Shengding Hu, Yulin Chen, Chi-Min Chan, Weize Chen, and 1 others. 2023.
\newblock Parameter-efficient fine-tuning of large-scale pre-trained language models.
\newblock \emph{Nature Machine Intelligence}, 5(3):220--235.

\bibitem[{Dong et~al.(2018)Dong, Liao, Pang, Su, Zhu, Hu, and Li}]{dong2018boosting}
Yinpeng Dong, Fangzhou Liao, Tianyu Pang, Hang Su, Jun Zhu, Xiaolin Hu, and Jianguo Li. 2018.
\newblock Boosting adversarial attacks with momentum.
\newblock In \emph{Computer Vision and Pattern Recognition Conference (CVPR)}, pages 9185--9193.

\bibitem[{Fang et~al.(2023)Fang, Wang, Xie, Sun, Wu, Wang, Huang, Wang, and Cao}]{fang2023eva}
Yuxin Fang, Wen Wang, Binhui Xie, Quan Sun, Ledell Wu, Xinggang Wang, Tiejun Huang, Xinlong Wang, and Yue Cao. 2023.
\newblock Eva: Exploring the limits of masked visual representation learning at scale.
\newblock In \emph{Computer Vision and Pattern Recognition Conference (CVPR)}.

\bibitem[{Goodfellow et~al.(2015)Goodfellow, Shlens, and Szegedy}]{goodfellow2015explaining}
Ian~J Goodfellow, Jonathon Shlens, and Christian Szegedy. 2015.
\newblock Explaining and harnessing adversarial examples.
\newblock \emph{stat}, 1050:20.

\bibitem[{Li et~al.(2023)Li, Li, Savarese, and Hoi}]{li2023blip}
Junnan Li, Dongxu Li, Silvio Savarese, and Steven Hoi. 2023.
\newblock Blip-2: Bootstrapping language-image pre-training with frozen image encoders and large language models.
\newblock In \emph{International Conference on Machine Learning (ICML)}, pages 19730--19742.

\bibitem[{Lin et~al.(2014)Lin, Maire, Belongie, Hays, Perona, Ramanan, Doll{\'a}r, and Zitnick}]{lin2014microsoft}
Tsung-Yi Lin, Michael Maire, Serge Belongie, James Hays, Pietro Perona, Deva Ramanan, Piotr Doll{\'a}r, and C~Lawrence Zitnick. 2014.
\newblock Microsoft coco: Common objects in context.
\newblock In \emph{Computer vision--ECCV 2014: 13th European conference, zurich, Switzerland, September 6-12, 2014, proceedings, part v 13}, pages 740--755. Springer.

\bibitem[{Liu et~al.(2024)Liu, Li, Li, Li, Zhang, Shen, and Lee}]{liu2024llavanext}
Haotian Liu, Chunyuan Li, Yuheng Li, Bo~Li, Yuanhan Zhang, Sheng Shen, and Yong~Jae Lee. 2024.
\newblock \href {https://llava-vl.github.io/blog/2024-01-30-llava-next/} {Llava-next: Improved reasoning, ocr, and world knowledge}.

\bibitem[{Liu et~al.(2023{\natexlab{a}})Liu, Li, Wu, and Lee}]{liu2024visual}
Haotian Liu, Chunyuan Li, Qingyang Wu, and Yong~Jae Lee. 2023{\natexlab{a}}.
\newblock \href {https://proceedings.neurips.cc/paper_files/paper/2023/file/6dcf277ea32ce3288914faf369fe6de0-Paper-Conference.pdf} {Visual instruction tuning}.
\newblock In \emph{Advances in Neural Information Processing Systems (NeurIPS)}, volume~36, pages 34892--34916.

\bibitem[{Liu et~al.(2023{\natexlab{b}})Liu, Deng, Xu, Li, Zheng, Zhang, Zhao, Zhang, and Liu}]{liu2023jailbreaking}
Yi~Liu, Gelei Deng, Zhengzi Xu, Yuekang Li, Yaowen Zheng, Ying Zhang, Lida Zhao, Tianwei Zhang, and Yang Liu. 2023{\natexlab{b}}.
\newblock Jailbreaking chatgpt via prompt engineering: An empirical study.
\newblock \emph{arXiv preprint arXiv:2305.13860}.

\bibitem[{Madry et~al.(2018)Madry, Makelov, Schmidt, Tsipras, and Vladu}]{madry2018towards}
Aleksander Madry, Aleksandar Makelov, Ludwig Schmidt, Dimitris Tsipras, and Adrian Vladu. 2018.
\newblock Towards deep learning models resistant to adversarial attacks.
\newblock In \emph{International Conference on Learning Representations (ICLR)}.

\bibitem[{Nguyen et~al.(2015)Nguyen, Yosinski, and Clune}]{nguyen2015deep}
Anh Nguyen, Jason Yosinski, and Jeff Clune. 2015.
\newblock Deep neural networks are easily fooled: High confidence predictions for unrecognizable images.
\newblock In \emph{Computer Vision and Pattern Recognition Conference (CVPR)}.

\bibitem[{Radford et~al.(2021)Radford, Kim, Hallacy, Ramesh, Goh, Agarwal, Sastry, Askell, Mishkin, Clark et~al.}]{radford2021learning}
Alec Radford, Jong~Wook Kim, Chris Hallacy, Aditya Ramesh, Gabriel Goh, Sandhini Agarwal, Girish Sastry, Amanda Askell, Pamela Mishkin, Jack Clark, and 1 others. 2021.
\newblock Learning transferable visual models from natural language supervision.
\newblock In \emph{International Conference on Machine Learning (ICML)}, pages 8748--8763.

\bibitem[{Ren et~al.(2024)Ren, Yao, Li, Sun, and Hou}]{ren2024timechat}
Shuhuai Ren, Linli Yao, Shicheng Li, Xu~Sun, and Lu~Hou. 2024.
\newblock Timechat: A time-sensitive multimodal large language model for long video understanding.
\newblock In \emph{Proceedings of the IEEE/CVF Conference on Computer Vision and Pattern Recognition}, pages 14313--14323.

\bibitem[{Shayegani et~al.(2023)Shayegani, Dong, and Abu-Ghazaleh}]{shayegani2023jailbreak}
Erfan Shayegani, Yue Dong, and Nael Abu-Ghazaleh. 2023.
\newblock Jailbreak in pieces: Compositional adversarial attacks on multi-modal language models.
\newblock \emph{arXiv preprint arXiv:2307.14539}.

\bibitem[{Shayegani et~al.(2024)Shayegani, Dong, and Abu-Ghazaleh}]{shayegani2023survey}
Erfan Shayegani, Yue Dong, and Nael Abu-Ghazaleh. 2024.
\newblock \href {https://openreview.net/forum?id=plmBsXHxgR} {Jailbreak in pieces: Compositional adversarial attacks on multi-modal language models}.
\newblock In \emph{International Conference on Learning Representations (ICLR)}.

\bibitem[{Shi et~al.(2024)Shi, Liu, Wang, Liao, Radhakrishnan, Huang, Yin, Sapra, Yacoob, Shi et~al.}]{shi2024eagle}
Min Shi, Fuxiao Liu, Shihao Wang, Shijia Liao, Subhashree Radhakrishnan, De-An Huang, Hongxu Yin, Karan Sapra, Yaser Yacoob, Humphrey Shi, and 1 others. 2024.
\newblock Eagle: Exploring the design space for multimodal llms with mixture of encoders.
\newblock \emph{arXiv preprint arXiv:2408.15998}.

\bibitem[{Szegedy et~al.(2014)Szegedy, Zaremba, Sutskever, Bruna, Erhan, Goodfellow, and Fergus}]{szegedy2014intriguing}
Christian Szegedy, Wojciech Zaremba, Ilya Sutskever, Joan Bruna, Dumitru Erhan, Ian Goodfellow, and Rob Fergus. 2014.
\newblock Intriguing properties of neural networks.
\newblock In \emph{International Conference on Learning Representations (ICLR)}.

\bibitem[{Touvron et~al.(2023)Touvron, Lavril, Izacard, Martinet, Lachaux, Lacroix, Rozi{\`e}re, Goyal, Hambro, Azhar et~al.}]{touvron2023llama}
Hugo Touvron, Thibaut Lavril, Gautier Izacard, Xavier Martinet, Marie-Anne Lachaux, Timoth{\'e}e Lacroix, Baptiste Rozi{\`e}re, Naman Goyal, Eric Hambro, Faisal Azhar, and 1 others. 2023.
\newblock Llama: Open and efficient foundation language models.
\newblock \emph{arXiv preprint arXiv:2302.13971}.

\bibitem[{Vedantam et~al.(2015)Vedantam, Lawrence~Zitnick, and Parikh}]{vedantam2015cider}
Ramakrishna Vedantam, C~Lawrence~Zitnick, and Devi Parikh. 2015.
\newblock Cider: Consensus-based image description evaluation.
\newblock In \emph{Proceedings of the IEEE conference on computer vision and pattern recognition}, pages 4566--4575.

\bibitem[{Wang et~al.(2024)Wang, Bai, Tan, Wang, Fan, Bai, Chen, Liu, Wang, Ge et~al.}]{wang2024qwen2}
Peng Wang, Shuai Bai, Sinan Tan, Shijie Wang, Zhihao Fan, Jinze Bai, Keqin Chen, Xuejing Liu, Jialin Wang, Wenbin Ge, and 1 others. 2024.
\newblock Qwen2-vl: Enhancing vision-language model's perception of the world at any resolution.
\newblock \emph{arXiv preprint arXiv:2409.12191}.

\bibitem[{Wang et~al.(2023)Wang, Ji, Ma, Li, and Wang}]{wang2023instructta}
Xunguang Wang, Zhenlan Ji, Pingchuan Ma, Zongjie Li, and Shuai Wang. 2023.
\newblock Instructta: Instruction-tuned targeted attack for large vision-language models.
\newblock \emph{arXiv preprint arXiv:2312.01886}.

\bibitem[{Xie et~al.(2019)Xie, Zhang, Zhou, Bai, Wang, Ren, and Yuille}]{xie2019improving}
Cihang Xie, Zhishuai Zhang, Yuyin Zhou, Song Bai, Jianyu Wang, Zhou Ren, and Alan~L Yuille. 2019.
\newblock Improving transferability of adversarial examples with input diversity.
\newblock In \emph{Computer Vision and Pattern Recognition Conference (CVPR)}, pages 2730--2739.

\bibitem[{Yao et~al.(2024)Yao, Li, Ren, Wang, Liu, Sun, and Hou}]{yao2024deco}
Linli Yao, Lei Li, Shuhuai Ren, Lean Wang, Yuanxin Liu, Xu~Sun, and Lu~Hou. 2024.
\newblock Deco: Decoupling token compression from semantic abstraction in multimodal large language models.
\newblock \emph{arXiv preprint arXiv:2405.20985}.

\bibitem[{Yin et~al.(2023)Yin, Ye, Zhang, Du, Zhu, Liu, Chen, Wang, and Ma}]{yin2024vlattack}
Ziyi Yin, Muchao Ye, Tianrong Zhang, Tianyu Du, Jinguo Zhu, Han Liu, Jinghui Chen, Ting Wang, and Fenglong Ma. 2023.
\newblock \href {https://openreview.net/forum?id=qBAED3u1XZ} {{VLATTACK}: Multimodal adversarial attacks on vision-language tasks via pre-trained models}.
\newblock In \emph{Advances in Neural Information Processing Systems (NeurIPS)}.

\bibitem[{Zhang et~al.(2022)Zhang, Roller, Goyal, Artetxe, Chen, Chen, Dewan, Diab, Li, Lin et~al.}]{zhang2022opt}
Susan Zhang, Stephen Roller, Naman Goyal, Mikel Artetxe, Moya Chen, Shuohui Chen, Christopher Dewan, Mona Diab, Xian Li, Xi~Victoria Lin, and 1 others. 2022.
\newblock Opt: Open pre-trained transformer language models.
\newblock \emph{arXiv preprint arXiv:2205.01068}.

\bibitem[{Zhang et~al.(2025)Zhang, Xie, Chen, Sun, Kang, and Wang}]{zhang2025qava}
Yudong Zhang, Ruobing Xie, Jiansheng Chen, Xingwu Sun, Zhanhui Kang, and Yu~Wang. 2025.
\newblock Qava: Query-agnostic visual attack to large vision-language models.
\newblock In \emph{Proceedings of the 2025 Conference of the Nations of the Americas Chapter of the Association for Computational Linguistics: Human Language Technologies (Volume 1: Long Papers)}, pages 10205--10218.

\bibitem[{Zhu et~al.(2024)Zhu, Chen, Shen, Li, and Elhoseiny}]{zhu2023minigpt}
Deyao Zhu, Jun Chen, Xiaoqian Shen, Xiang Li, and Mohamed Elhoseiny. 2024.
\newblock \href {https://openreview.net/forum?id=1tZbq88f27} {Mini{GPT}-4: Enhancing vision-language understanding with advanced large language models}.
\newblock In \emph{International Conference on Learning Representations (ICLR)}.

\bibitem[{Zou et~al.(2023)Zou, Wang, Carlini, Nasr, Kolter, and Fredrikson}]{zou2023universal}
Andy Zou, Zifan Wang, Nicholas Carlini, Milad Nasr, J~Zico Kolter, and Matt Fredrikson. 2023.
\newblock Universal and transferable adversarial attacks on aligned language models.
\newblock \emph{arXiv preprint arXiv:2307.15043}.

\end{thebibliography}

\clearpage
\appendix
\onecolumn

\section{Visualized Examples of Attacks}
\label{sec:demos}

\begin{figure*}[ht]
    \centering
    \begin{minipage}{0.35\linewidth}
        \begin{itemize}
            \item[$\bullet$] Instruction: 
            \item[] What kind of bear is in the photo?
            \item[$\bullet$] Answer (clean):
            \item[] Polar.
            \item[$\bullet$] Answer (adversarial):
            \item[] Teddy.
            \item[] ------------------------------ 
        \end{itemize}        
    \end{minipage}
    \hfill
    \begin{minipage}{0.64\linewidth}
        \centering
        \begin{subfigure}{0.3\linewidth}
            \includegraphics[width=\linewidth]{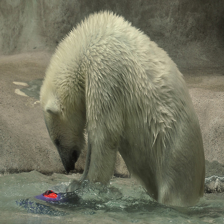}
        \end{subfigure}
        \begin{subfigure}{0.3\linewidth}
            \includegraphics[width=\linewidth]{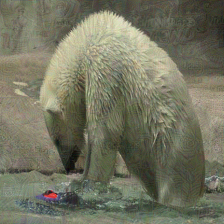}
        \end{subfigure}
        \begin{subfigure}{0.3\linewidth}
            \includegraphics[width=\linewidth]{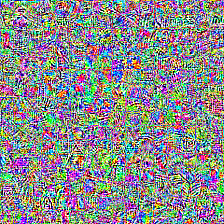}
        \end{subfigure}
    \end{minipage}
    \begin{minipage}{0.35\linewidth}
        \begin{itemize}
            \item[$\bullet$] Instruction: 
            \item[] Is the girl moving?
            \item[$\bullet$] Answer (clean):
            \item[] Yes.
            \item[$\bullet$] Answer (adversarial):
            \item[] No.
            \item[] ------------------------------ 
        \end{itemize}        
    \end{minipage}
    \hfill
    \begin{minipage}{0.64\linewidth}
        \centering
        \begin{subfigure}{0.3\linewidth}
            \includegraphics[width=\linewidth]{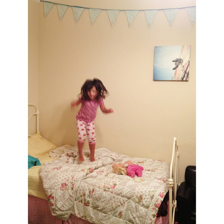}
        \end{subfigure}
        \begin{subfigure}{0.3\linewidth}
            \includegraphics[width=\linewidth]{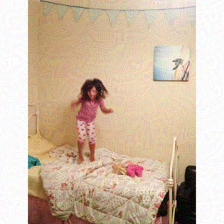}
        \end{subfigure}
        \begin{subfigure}{0.3\linewidth}
            \includegraphics[width=\linewidth]{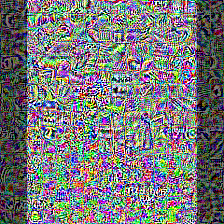}
        \end{subfigure}
    \end{minipage}
    \begin{minipage}{0.35\linewidth}
        \begin{itemize}
            \item[$\bullet$] Instruction: 
            \item[] Is this a kitchen?
            \item[$\bullet$] Answer (clean):
            \item[] No.
            \item[$\bullet$] Answer (adversarial):
            \item[] Yes.
            \item[] ------------------------------ 
        \end{itemize}        
    \end{minipage}
    \hfill
    \begin{minipage}{0.64\linewidth}
        \centering
        \begin{subfigure}{0.3\linewidth}
            \includegraphics[width=\linewidth]{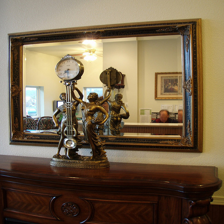}
        \end{subfigure}
        \begin{subfigure}{0.3\linewidth}
            \includegraphics[width=\linewidth]{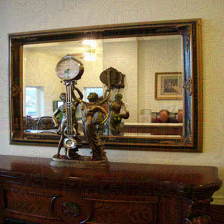}
        \end{subfigure}
        \begin{subfigure}{0.3\linewidth}
            \includegraphics[width=\linewidth]{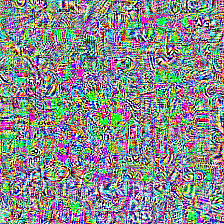}
        \end{subfigure}
    \end{minipage}
    \begin{minipage}{0.35\linewidth}
        \begin{itemize}
            \item[$\bullet$] Instruction: 
            \item[] What is on the sign on the building?
            \item[$\bullet$] Answer (clean):
            \item[] Google.
            \item[$\bullet$] Answer (adversarial):
            \item[] Stop.
            \item[] ------------------------------ 
        \end{itemize}        
    \end{minipage}
    \hfill
    \begin{minipage}{0.64\linewidth}
        \centering
        \begin{subfigure}{0.3\linewidth}
            \includegraphics[width=\linewidth]{figure-supp/COCO_val2014_000000145073_ori.png}  
        \end{subfigure}
        \begin{subfigure}{0.3\linewidth}
            \includegraphics[width=\linewidth]{figure-supp/COCO_val2014_000000145073_adv.png}   
        \end{subfigure}
        \begin{subfigure}{0.3\linewidth}
            \includegraphics[width=\linewidth]{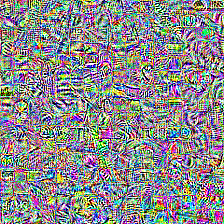}   
        \end{subfigure}
    \end{minipage}
    \caption{Some adversarial examples of attacks on InstructBLIP-Vicuna-7B.}
    \label{fig:figure5}
\end{figure*}

\begin{figure*}[ht]
    \centering
    \begin{minipage}{0.35\linewidth}
        \begin{itemize}
            \item[$\bullet$] Instruction: 
            \item[] How many children?
            \item[$\bullet$] Answer (clean):
            \item[] 2.
            \item[$\bullet$] Answer (adversarial):
            \item[] 0.
            \item[] ------------------------------ 
        \end{itemize}        
    \end{minipage}
    \hfill
    \begin{minipage}{0.64\linewidth}
        \centering
        \begin{subfigure}{0.3\linewidth}
            \includegraphics[width=\linewidth]{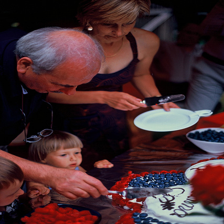}
        \end{subfigure}
        \begin{subfigure}{0.3\linewidth}
            \includegraphics[width=\linewidth]{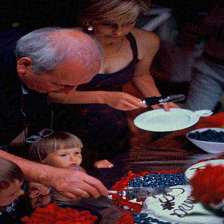}
        \end{subfigure}
        \begin{subfigure}{0.3\linewidth}
            \includegraphics[width=\linewidth]{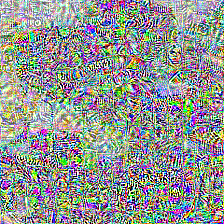}
        \end{subfigure}
    \end{minipage}
    \begin{minipage}{0.35\linewidth}
        \begin{itemize}
            \item[$\bullet$] Instruction: 
            \item[] Are they at the beach?
            \item[$\bullet$] Answer (clean):
            \item[] Yes.
            \item[$\bullet$] Answer (adversarial):
            \item[] No.
            \item[] ------------------------------ 
        \end{itemize}        
    \end{minipage}
    \hfill
    \begin{minipage}{0.64\linewidth}
        \centering
        \begin{subfigure}{0.3\linewidth}
            \includegraphics[width=\linewidth]{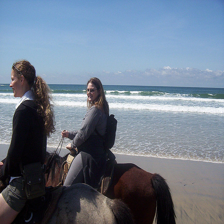}
        \end{subfigure}
        \begin{subfigure}{0.3\linewidth}
            \includegraphics[width=\linewidth]{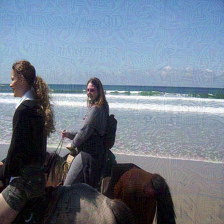}
        \end{subfigure}
        \begin{subfigure}{0.3\linewidth}
            \includegraphics[width=\linewidth]{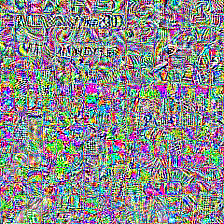}
        \end{subfigure}
    \end{minipage}
    \begin{minipage}{0.35\linewidth}
        \begin{itemize}
            \item[$\bullet$] Instruction: 
            \item[] What sport is this?
            \item[$\bullet$] Answer (clean):
            \item[] Tennis.
            \item[$\bullet$] Answer (adversarial):
            \item[] Skateboarding.
            \item[] ------------------------------ 
        \end{itemize}        
    \end{minipage}
    \hfill
    \begin{minipage}{0.64\linewidth}
        \centering
        \begin{subfigure}{0.3\linewidth}
            \includegraphics[width=\linewidth]{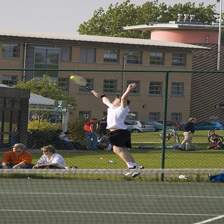}
        \end{subfigure}
        \begin{subfigure}{0.3\linewidth}
            \includegraphics[width=\linewidth]{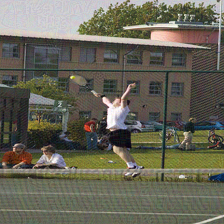}
        \end{subfigure}
        \begin{subfigure}{0.3\linewidth}
            \includegraphics[width=\linewidth]{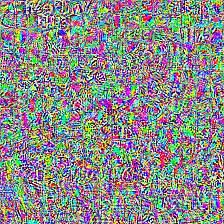}
        \end{subfigure}
    \end{minipage}
    \begin{minipage}{0.35\linewidth}
        \begin{itemize}
            \item[$\bullet$] Instruction: 
            \item[] What is the name on the front of the truck
            \item[$\bullet$] Answer (clean):
            \item[] Dodge.
            \item[$\bullet$] Answer (adversarial):
            \item[] Force.
            \item[] ------------------------------ 
        \end{itemize}        
    \end{minipage}
    \hfill
    \begin{minipage}{0.64\linewidth}
        \centering
        \begin{subfigure}{0.3\linewidth}
            \includegraphics[width=\linewidth]{figure-supp/COCO_val2014_000000124756_ori.png}
        \end{subfigure}
        \begin{subfigure}{0.3\linewidth}
            \includegraphics[width=\linewidth]{figure-supp/COCO_val2014_000000124756_adv.png}
        \end{subfigure}
        \begin{subfigure}{0.3\linewidth}
            \includegraphics[width=\linewidth]{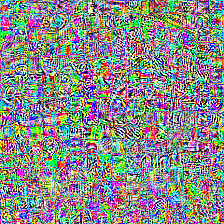}
        \end{subfigure}
    \end{minipage}
    \caption{Some adversarial examples of attacks on InstructBLIP-Vicuna-7B.}
    \label{fig:figure6}
\end{figure*}

We present some visualized examples from VQA v2 2000 and InstructBLIP-Vicuna-7B in \cref{fig:figure5} and \cref{fig:figure6}.

Some things to note are the following.

\begin{itemize}
    \item The experimental setting involves $\mathcal{L}_\text{TCP}(\beta=0.2, N=2)$ combined with a PGD attack.
    % \item The casing of answers (\eg, Yes or yes, No or no) does not impact the results, as casing is standardized in the official evaluation code for VQA v2. 
    \item The adversarial images can target to various versions of InstructBLIP, and we only show results for InstructBLIP(Vicuna-7B).
    \item For each row, the left image is the clean image, the center image is the adversarial image, and the right image is the adversarial perturbation magnified by a factor of 16 (\ie, when  perturbation limitation $\epsilon=8$, $-\epsilon=-8\to0$ and $\epsilon=8\to255$). The adversarial image is obtained from a clean image plus an adversarial perturbation.
\end{itemize} 

These examples demonstrate that the attack successfully misdirects the LVLM even under the gray-box setting, highlighting security vulnerabilities in compressed projectors.

\end{document}